\documentclass[%
 aip, apl
 amsmath,amssymb,
 reprint,%
]{revtex4-1}

\usepackage{graphicx}
\usepackage{dcolumn}
\usepackage{bm}
\usepackage[colorlinks=true]{hyperref}
\usepackage[dvipsnames]{xcolor}

\colorlet{mylinkcolor}{RoyalPurple}
\colorlet{mycitecolor}{RoyalPurple}
\colorlet{myurlcolor}{RoyalPurple}
\hypersetup{
	linkcolor  = mylinkcolor,
	citecolor  = mycitecolor,
	urlcolor   = myurlcolor,
	colorlinks = true,
	breaklinks = true
}
\usepackage[utf8]{inputenc}
\usepackage[T1]{fontenc}
\usepackage{mathptmx}
\usepackage{siunitx}
\usepackage{etoolbox}

\newcommand{\E}{\mathcal{E}}
\usepackage{todonotes}
\usepackage{nicefrac}
\usepackage{lineno}
\usepackage[english]{babel}
\usepackage[dvipsnames]{xcolor}
\makeatletter
\def\@email#1#2{%
 \endgroup
 \patchcmd{\titleblock@produce}
  {\frontmatter@RRAPformat}
  {\frontmatter@RRAPformat{\produce@RRAP{*#1\href{mailto:#2}{#2}}}\frontmatter@RRAPformat}
  {}{}
}%
\makeatother
\begin{document}


\title{Coherent terahertz field tomographic imaging in warm Rydberg vapors} 


\author{Jan Nowosielski}
\affiliation{Faculty of Physics, University of Warsaw, L. Pasteura 5, 02-093 Warsaw, Poland}
\affiliation{Centre for Quantum Optical Technologies, Centre of New Technologies, University of Warsaw, S. Banacha 2c, 02-097 Warsaw, Poland}
\email[]{j.nowosielski@cent.uw.edu.pl}

\author{Marcin Jastrzębski}
\affiliation{Faculty of Physics, University of Warsaw, L. Pasteura 5, 02-093 Warsaw, Poland}
\affiliation{Centre for Quantum Optical Technologies, Centre of New Technologies, University of Warsaw, S. Banacha 2c, 02-097 Warsaw, Poland}
\author{Wojciech Wasilewski}
\affiliation{Faculty of Physics, University of Warsaw, L. Pasteura 5, 02-093 Warsaw, Poland}
\affiliation{Centre for Quantum Optical Technologies, Centre of New Technologies, University of Warsaw, S. Banacha 2c, 02-097 Warsaw, Poland}
\author{Mateusz Mazelanik}
\affiliation{Centre for Quantum Optical Technologies, Centre of New Technologies, University of Warsaw, S. Banacha 2c, 02-097 Warsaw, Poland}
\author{Michał Parniak}
\affiliation{Faculty of Physics, University of Warsaw, L. Pasteura 5, 02-093 Warsaw, Poland}
\affiliation{Centre for Quantum Optical Technologies, Centre of New Technologies, University of Warsaw, S. Banacha 2c, 02-097 Warsaw, Poland}


\date{\today}

\begin{abstract}
Rydberg atom-based sensors have emerged as highly sensitive tools for terahertz (THz) metrology, yet most current imaging techniques discard crucial phase information. In this Letter, we present a coherent THz-to-optical conversion scheme in warm Rb vapor that enables complex-amplitude field imaging. By manipulating the phase-matching conditions via an adjustable interference pattern of optical probe beams, we demonstrate the ability to perform tomographic reconstruction of the THz field distribution. We experimentally validate the spatial resolution and phase-sensitivity of the system by resolving sub-centimeter features and identifying incident angles of arrival. Our results establish a robust framework for phase-resolved THz imaging and holography using atomic vapors at room temperature.\end{abstract}


\maketitle 

Terahertz (THz) radiation has recently attracted considerable attention, driven by rapid advances in photonics and materials science. This growing interest within the scientific and engineering communities has translated into an expanding range of applications, including security screening \cite{Federici_2005, Appleby2007, Davies2008}, wireless communications \cite{Liu2024, Yuan2023a}, and, in particular, field imaging \cite{KleineOstmann2001, Hu_1995}. The latter has become an essential tool for in vivo imaging of biological tissues \cite{Loeffler2001, Oh2013, Oh2014}, characterisation of nanoscale structures \cite{Chen2003, Huber2008, Hillenbrand2025}, and related high-resolution material studies.

In recent years, numerous THz imaging implementations have been proposed. Among them, Rydberg atom–based sensors have emerged as especially promising due to their high sensitivity to radio-frequency or THz fields and their capability for comprehensive field characterisation, including SI-traceable intensity \cite{Sedlacek2012}, polarisation \cite{Wang2023, Elgee2024}, and angle of arrival \cite{Talashila2025, Robinson2021, Schlossberger2024a}. Imaging with such platforms has already been demonstrated, primarily through measurements of atomic fluorescence \cite{Schlossberger2025b, Downes2023, Wade2016, Pati2026} or laser transmission detected using either a photodiode \cite{Holloway2014} or a CCD camera \cite{Fan2014}. Despite substantial progress, these techniques remain limited to intensity-only imaging of the THz field, thereby discarding crucial phase information.

In this work, we present a THz field imaging approach that addresses this limitation by exploiting a THz-to-optical conversion process, enabling retrieval of the complex amplitude of the incident field, to which we further refer to as coherent imaging. We demonstrate the feasibility of the proposed method by reconstructing the field distribution in two representative experimental scenarios.
 
\begin{figure*}
    \centering
    \includegraphics[width=\linewidth]{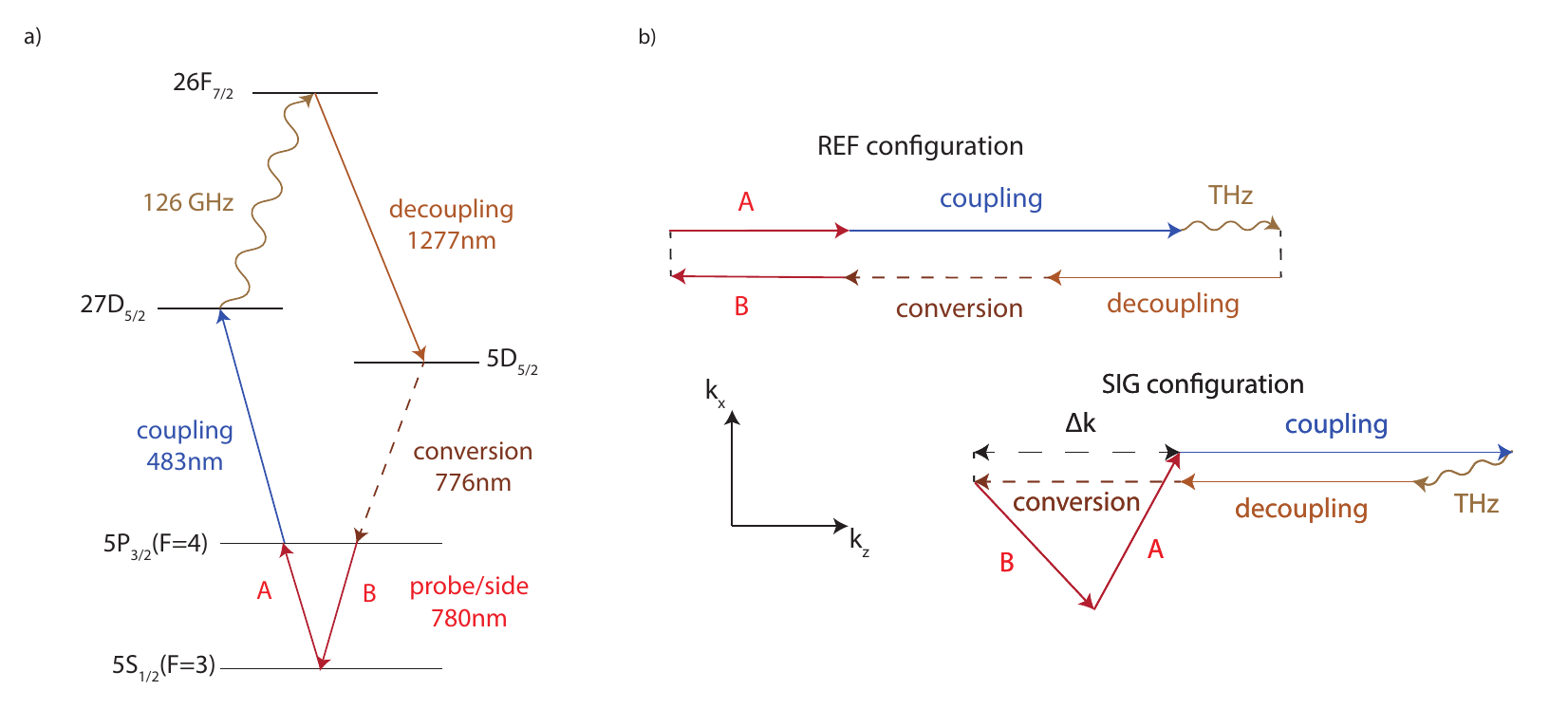}
    \caption{a) Energy level scheme used in the experiment. b) Schematic depiction of two wavevector configurations considered in the experiment. For the sake of example, the wavevectors are only depicted in the two-dimensional space, where $k_z$ is the wavevector component along the optical axis, and $k_x$ is the component perpendicular to it. To improve the clarity of the scheme, the length of the wavevectors is not up to scale, and the relation between them is not maintained. Additionally, the wavevectors in the REF configurations are shifted in $k_x$.}
    \label{fig:configs}
\end{figure*}
The experiment revolves around the process called THz-to-optical conversion, which is a specific case of the multiwave-mixing process utilising a nonlinear atomic medium in the configuration shown in Fig. \ref{fig:configs}a. In this case, the oncoming THz field is being absorbed by the atomic medium and coherently converted into visible light. During the experiment, the wave-mixing process occurs when the atomic medium interacts with the probe, coupling and decoupling lasers, allowing for the conversion of the THz field into the conversion signal in the optical domain.

All the fields interacting with the medium can be defined as:
\begin{equation}
    \E_i(\vec r, t) = A_i(\vec r) \exp(-i\omega_i t + i \vec k_i\cdot \vec r) 
\end{equation}
where $A_i$, $\omega_i$, and $\vec {k}_i$ are respectively the complex amplitude, frequency and wavevector of the particular field. For the sake of a clearer picture of the working principle, all the fields interacting with the atomic medium can be assumed to be plane waves. Additionally, under the assumption that the converted field is being emitted along the optical z-axis, its amplitude can be written as:
\begin{equation}\label{eq:optical_full}
    A_o = \int_0^L \text{d}z \exp(-ik_o z)
\frac{\omega_o^2}{2ic^2k_o \epsilon_0} \mathcal{P}_{o}
\end{equation}
where $\epsilon_0$ is the vacuum permitivity, and $L$ is the length of atomic medium. The $\mathcal{P}_o$ is the nonlinear polarisation induced by all the fields interacting with the atomic medium:
\begin{equation}
    \mathcal{P}_o = \chi^{(6)}\E_A\E_c\E_{THz}\E^*_d\E^*_B
\end{equation}
where $\chi^{(6)}$ is nonlinear susceptibility, $\E_A$ and $\E_B$ are the absorbed and emitted probe fields, $\E_c$ is the coupling field, and $\E_d$ is the decoupling field. Assuming additionally that the coupling and decoupling fields are constant along the atomic medium, Eq. \ref{eq:optical_full} can be rewritten as:
\begin{equation}\label{eq:optical_short}
    A_o = K\int_0^L \text{d}z\,\, C(z)A_{THz}(z)\exp(i\delta k z)
\end{equation}
where $K = \frac{\omega_o^2\chi^{(6)}}{2ic^2k_o \epsilon_0}A_cA_d$, and $\delta k$ is the conversion wavevector mismatch defined as:
\begin{equation}
    \delta k = k_c + k^{z}_{THz} - k_d - k_o
\end{equation}
As the dimensions of the atomic medium perpendicular to the optical axis are negligible, wavevector mismatch in their direction can be omitted when considering the phase-matching, thus allowing only for the consideration of the projections of the wavevectors onto the z-axis.
The $C(z)$ factor represents the impact of the A and B fields on the conversion signal and is given as:
\begin{equation}
    C(z) = \E_A\E^*_B = A_A A^*_B\exp(i\Delta k z)
\end{equation}
where $\Delta k  = k_A - k_B$ is the wavevector difference between A and B fields along the z-axis. For a long enough atomic medium, the conversion occurs only when the proper phase-matching condition is met, which can be written as:
\begin{equation}
    \Delta k + \delta k = 0
\end{equation}

\begin{figure*}
    \centering
    \includegraphics[width=\linewidth]{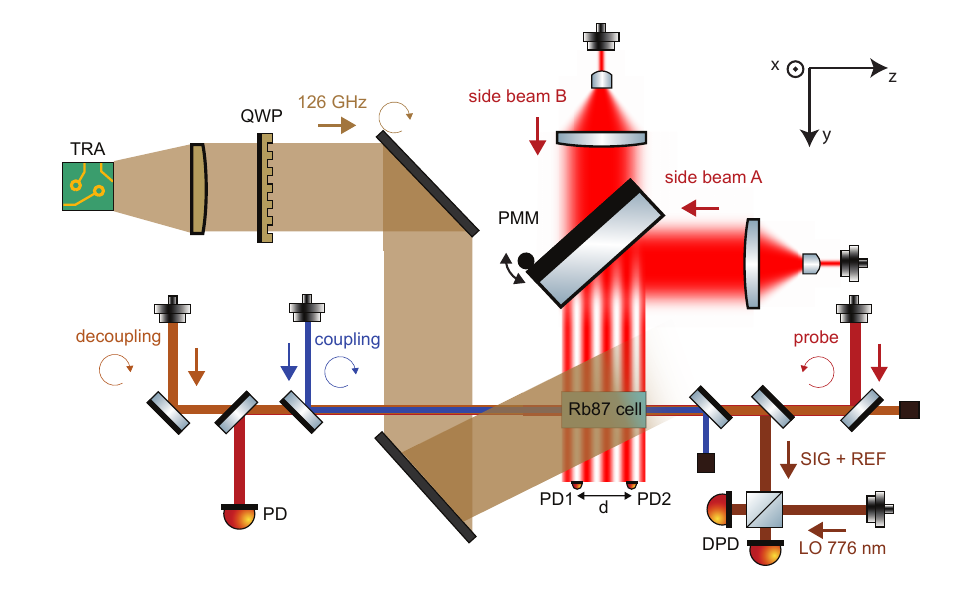}
    \caption{Schematic depiction of the experimental setup. QWP - quarter waveplate, PD - photodiode, DPD - differential photodiode, PMM - piezo-mounted mirror, LO - local oscillator, SIG - signal, REF - reference.}
    \label{fig:setup}
\end{figure*}

During the experiment, two distinct field configurations are considered, namely reference (REF) and signal (SIG), shown in Fig. \ref{fig:configs}b. In the REF configuration, the A and B fields are the same laser beam, leading to $\E_A = \E_B$ and $\Delta k = 0$, thus also forcing $\delta k = 0$. Considering the experimental case, where coupling, decoupling and the conversion fields are propagating along the z-axis, the $\delta k = 0$ condition establishes the reference direction of the THz field along the optical axis with the wavevector $k_{THz}^{REF}$. In the SIG configuration, the coupling, decoupling and optical fields propagate in the same, fixed direction, as described before. On the other hand, A and B fields are now two separate side beams crossing inside the atomic medium at some given angle, yielding a non-zero value of $\Delta k$ and in turn creating an interference pattern as depicted in Fig.~\ref{fig:setup}. The value of $\Delta k$ is controlled by varying the angle of side beam A, thereby allowing phase-matching to different values of $\delta k$ and recovering the wavevector distribution of the oncoming THz field in terms of the mismatch $\delta k$ that can be interpreted as the difference between the reference THz wavevector and the projection of the signal wavevector onto the optical axis $k_{THz}$, given as:
\begin{equation}
    \delta k = k^{z}_{THz} - k_{THz}^{REF}.
\end{equation}
It is worth noting that the distribution of the THz field should only span values between $\Delta k = 0$ and $\Delta k = 2|k_{THz}|$ corresponding to the THz field co- and counterpropagating to the coupling beam, with the $|k_{THz}| = \frac{2\pi}{\lambda_{THz}}$ being the magnitude of THz field wavevector.

The experiment is built around a heated rectangular vapour cell containing warm Rb87 atoms. Its simplified scheme is depicted in Fig. \ref{fig:setup}a, and the energy level configuration used in the experiment is depicted in Fig. \ref{fig:configs}a. The basic working principle of the setup has already been presented in our previous works \cite{Borowka2024, Krokosz2025}. First, the atoms are excited from the ground $5^2S_{1/2}(F=2)$ using the 780 nm probe laser tuned to the D2 line to the $5^2P_{3/2}(F=3)$ state. Next, the atoms are excited to the Rydberg $27^2D_{5/2}$ via the \SI{483}{\nm} coupling laser. The atoms are further excited to the $26^2 F_{7/2}$ using the THz field at the frequency of $\SI{126}{\GHz}$, and subsequently deexcited to the state $5^2D_{5/2}$ via the \SI{1276}{\nm} decoupling laser. Lastly, the atoms deexcite back to the $5^2P_{3/2}(F=3)$ state by emitting the converted light at the wavelength of $\SI{776}{\nm}$. All the laser beams have a waist of around $w \approx \SI{250}{\micro \m}$. 

The \SI{780}{nm} side beams A and B, similarly to the probe beams, are tuned to the D2 line and illuminate the cell from its side, where they create an interference pattern. Both beams have a flat-top intensity profile with a width of about \SI{25}{\mm} and a height of about \SI{4}{\mm}, which is generated using Powell and a cylindrical lens. The beams are overlapped with each other using a piezo-driven D-shaped mirror, allowing for changing the angle between the beams in zx-plane, leading to controllable changes in the wavevector mismatch $\Delta k$.

In order to maximise the conversion signal, the cell is heated to about \SI{65}{\celsius} and the polarisation of all the interacting fields is circular. The powers of all the lasers taking part in the conversion process were also optimized to maximise the conversion signal, and were set to $P_c = \SI{150}{\milli\watt}$, $P_d = \SI{40}{\milli\watt}$, $P_p = \SI{20}{\micro\watt}$ and $P_A=P_B=\SI{23}{\milli\watt}$ for the coupling, decoupling, probe and side beams respectively.

The THz field detected in the experiment is generated using the TRA 120\_120\_45 chip manufactured by Indie Semiconductors. Such a field, after being emitted, passes through the collimating lens and quarter-waveplate. Both the lens and the quarter-waveplate are 3D-printed using the HIPS material \cite{Borowka2024a}. Next, the field is injected into the vapour cell using the metallic plates acting as THz mirrors. The average value of the Rabi frequency of the THz field inside the cell was determined using an auxiliary measurement of the electromagnetically induced transparency (EIT) spectrum \cite{Fleischhauer2005}, and is equal to about $\Omega_{THz}\approx 2\pi\cdot\SI{12,4}{\MHz}$.

The conversion signal $A_o$ is measured using heterodyne detection. The emitted field is overlapped at the differential photodiode with a stronger $\SI{776}{\nm}$ reference laser beam acting as a local oscillator (LO). The frequency of the local oscillator $\omega_{LO}$ is set to be detuned by about $\delta = \SI{3}{\MHz}$ from the frequency of the conversion signal $\omega_{SIG}$. The measured beat-note is then IQ demodulated and processed digitally using the StemLab 125-14 board.

\begin{figure*}
    \centering
    \includegraphics[width=\linewidth]{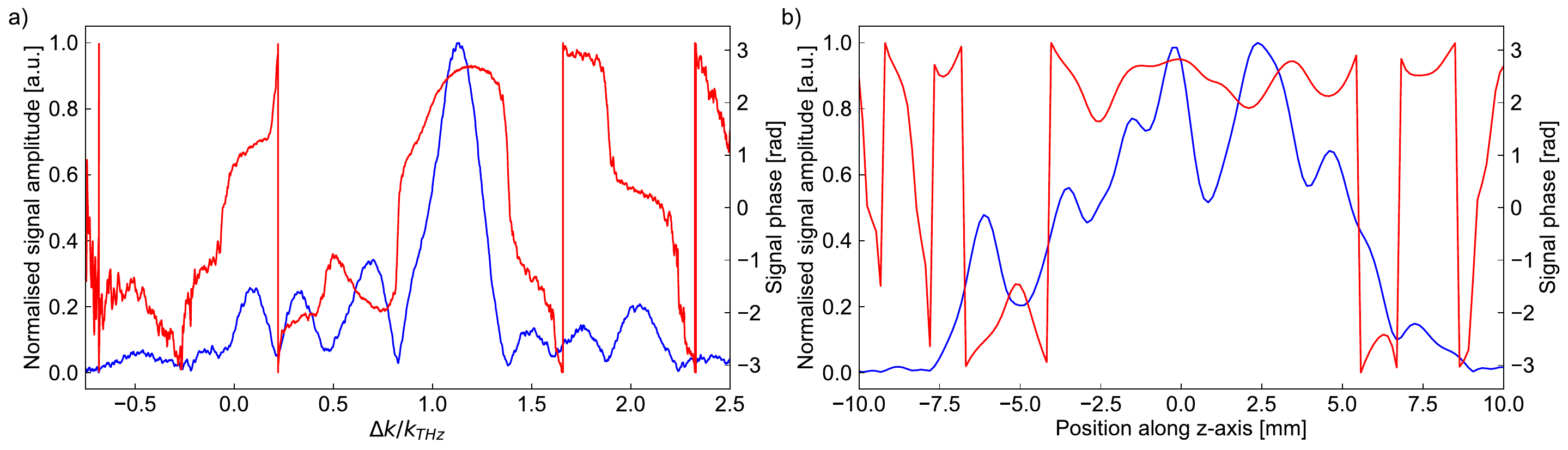}
    \caption{a) Amplitude of the SIG conversion signal (blue) and its phase (red) as a function of wavevector difference $\Delta k$. b) Amplitude (blue) and phase (red) of the spatial distribution of the THz field as a function of the position $z$ along the optical axis.}
    \label{fig:readout}
\end{figure*}
Due to finite linewidths of the lasers taking part in the conversion process, both signals $A_{REF}$ and $A_{SIG}$ are spread across approx. $\SI{0,5}{\MHz}$ of bandwidth, and thus have limited SNR. In fact, due to the lack of proper compensation for the Doppler effect in the SIG configuration, the conversion signal is so weak that it cannot be seen directly in the heterodyne detection. However, as the signals originate from the same laser fields, they have correlated noise, which we exploit to overcome that problem. For this, we measure the conversion signals $A_{REF}$ and $A_{SIG}$ from both the REF and SIG configurations simultaneously, allowing for the phase correction of $A_{SIG}$ and a significant increase in its visibility. To properly measure both signals, we separate them in the frequency domain by introducing the detuning between the side beams $\Delta\omega = \SI{2}{\MHz}$, leading to the SIG beat-note being shifted to the frequency $\delta+\Delta\omega$. The signal measured using the differential photodiode is then separately digitally demodulated on the frequencies $\Delta \omega$ and $\delta+\Delta\omega$, yielding $A_{REF}$ and $A_{SIG}$ signals. The phase correction is then achieved by electronically mixing the demodulated signals, resulting in the signal used in the final analysis. 


To accurately estimate the value of $\Delta k$ introduced by the side beams, two photodiodes, named PD1 and PD2, spaced by the distance $d \approx \SI{8}{\mm}$ are placed on the opposite side of the cell from the oncoming side beams. Both photodiodes measure the beat-note between side beams at the frequency of $\Delta\omega = \SI{2}{\MHz}$. The phases of both beat-note signals differ by the phase difference $\Delta\phi = \Delta k d + 2\pi n$, where $n$ is the number of full periods of the oscillation and cannot be determined directly from the photodiodes readout, requiring additional calibration.


The phase difference readout is calibrated using the Basler acA4096-11gm camera. For the sake of calibration, the side beams are redirected using the mirror at the camera placed at the same distance from the piezo-mounted mirror as the cell. Next, the $\Delta \omega$ is set to $\Delta\omega = 0$, and the interference pattern generated by the side beams is measured for values of the piezo driving voltage. By comparing the measured dependence of the interference pattern and the phase difference $\Delta\phi$, the proper calibration of the phase difference readout can be performed. 



The measurement of the side conversion signal SIG for different values of $\Delta k$ is performed using the heterodyne detection described above. The sweep of $\Delta k$ values is achieved by driving the piezo-mounted mirror with a triangular waveform with a frequency of $\SI{4}{\Hz}$. Such a measured signal is then demodulated and phase-corrected using the REF conversion signal and beat-note measured by the PD2 photodiode, leaving only the phase of the SIG signal. The corrected signal is then averaged over 100 waveforms to minimise the remaining measurement noise. The final effect of the procedure is the complex-amplitude conversion signal, whose exemplary amplitude and phase can be seen in Fig. \ref{fig:readout}a. For a better understanding of the presented THz field distribution, the wavevector mismatch $\Delta k$ has been scaled by the magnitude of the THz field wavevector $|k_{THz}|$. It can be noted that apart from the dominating term around $\Delta k = |k_{THz}|$, multiple smaller peaks are visible, which most likely can be attributed to the scattering and reflection from the aluminium cell holder. The dominating peak represents the main contributing wavevector of the THz field equal to about $\Delta k = \SI{1,13}{}|k_{THz}|$, which allows for a rough estimate of the angle-of-arrival of the field equal to about $\alpha \approx \SI{97}{\degree}$.

Having measured the complex-amplitude signal of the side conversion, it can then be Fourier transformed to recover the spatial distribution of the THz field along the optical axis. The exemplary spatial distribution and its phase corresponding to the previously presented conversion signal as a function of $\Delta k$ can be seen in Fig. \ref{fig:readout}b. Notably, the resolution of the spatial distribution is limited by the range of available values of $\Delta k$, in turn defined by the movement of the piezo stack used for the piezo mounted mirror. In the case of the described experimental setup, the spatial resolution is equal to about $\Delta z = \SI{0,7}{\mm}$.

\begin{figure*}
    \centering
    \includegraphics[width=\linewidth]{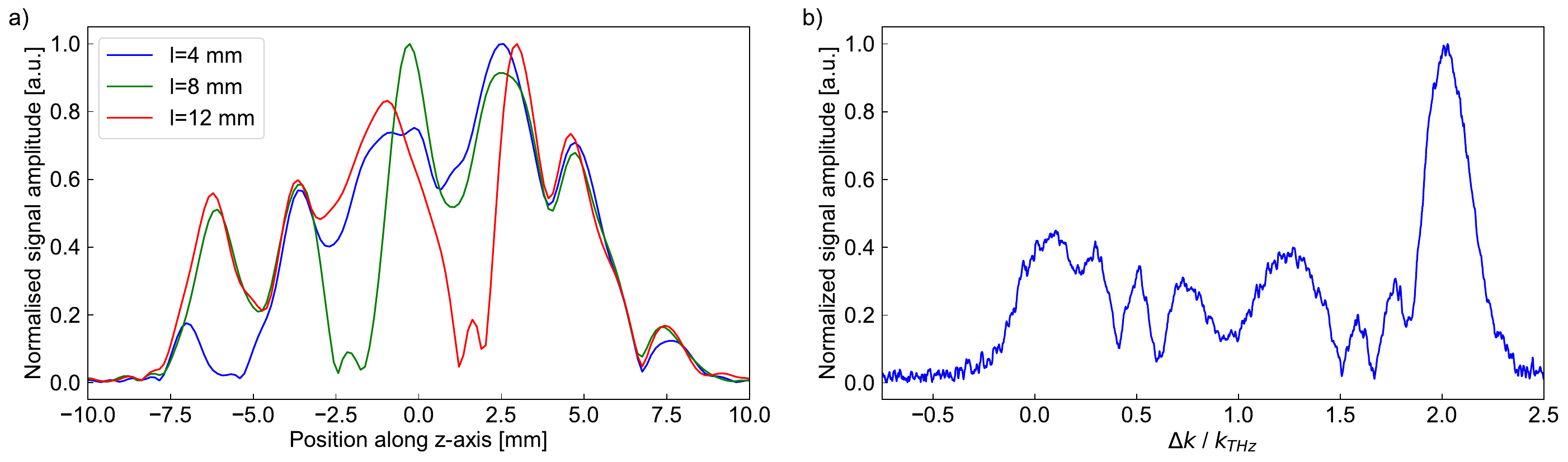}
    \caption{a) Amplitude of the spatial distribution of the THz field for different positions of the gap in the side beams, where $l$ is the distance from the edge of the cell. b) Amplitude of the conversion signal of the THz field injected using the waveguide through the back window of the cell as a function of the wavevector mismatch $\Delta k$.}
    \label{fig:dziura}
\end{figure*}

To further prove the coherence of the presented measurement method, part of the side beams were covered by the movable obstacle, leading to about $\SI{2}{\mm}$ gap in the side beams shining onto the cell. The position of the obstacle in the form of the plastic cylinder with a diameter of \SI{2}{\mm} was moved along the optical axis using the translation stage. The lack of side beams in a specific part of the cell leads to a lack of atomic excitation in that part, and thus, the THz field is not mapped. and the gap should be seen in the spatial distribution of the THz field. The side conversion signal as a function of $\Delta k$ was measured for different positions of the obstacle along the optical axis, and the spatial distribution in all the cases was calculated. The spatial distributions for three different positions of the obstacle can be seen in Fig. \ref{fig:dziura}. As expected, as the obstacle changed its position along the optical axis, the gap visible in all the signals changed its position accordingly.

Lastly, the incident THz field was altered by injecting it through the back cell window using a waveguide. The wavevector distribution in the described case can be seen in Fig. \ref{fig:dziura}b. Notably, the dominating peak lies close to $\Delta k = 2|k_{THz}|$, meaning that most of the THz field was counterpropagating to the reference direction. Additionally, two more distinct peaks can be seen, with the first lying around $\Delta k = 0$ corresponding to the reflection of the THz field from the window on the opposite side of the cell, and the second one lying around $\Delta k = \SI{1,3}{}|k_{THz}|$ corresponds to the THz field not coupled to the waveguide and propagating as in the previous configuration. The rest of the visible peaks can be attributed to the reflections and scattering inside the glass cell, as well as non-perfect coupling to the waveguide.

In this letter, we have presented a novel method of coherent THz field imaging, allowing for getting both amplitude and phase distribution of the oncoming field. We further proved the coherence of the measurement technique by altering the measured spatial distribution in an expected and controllable manner. While the results of the presented experiment pave the way for the new imaging technique, it still could be further improved. 

The described setup is currently limited by the two main factors: the resolution of the spatial distribution and the requirement for the REF signal. As the resolution of the spatial distribution is tied directly to the range of available values of $\Delta k$, and thus to the movement range of the piezo-mounted mirror, increasing the movement range would directly impact the resolution. Additionally, the current measurement scheme requires the REF signal and relies heavily on having a component of the THz field propagating in the reference direction. In principle, that problem could be mitigated by introducing a different way of laser phase noise correction, such as the generation of the reference laser using the multi-wave mixing process in the non-linear crystal \cite{Borowka2025}. 

While some limitations could still be mitigated in the future, we believe that the presented results already show the potential for the presented imaging technique. The described setup could additionally be expanded upon to perform the angle of arrival measurements of the oncoming THz field, as well as imaging of the entire 2D plane.

\begin{acknowledgments}
This research was funded in whole or in part by the National Science Centre, Poland, grants No.~2024/53/B/ST2/04040 and No.~2024/53/N/ST7/02730. The "Quantum Optical Technologies" (FENG.02.01-IP.05-0017/23) project is carried out within the Measure 2.1 International Research Agendas programme of the Foundation for Polish Science, co-financed by the European Union under the European Funds for Smart Economy 2021-2027 (FENG).
\end{acknowledgments}

\subsection*{Conflict of interest}
The authors have no conflicts to disclose.

\section*{Data availability}
Data has been deposited at Harvard Dataverse \cite{Nowosielski2026}


%
%

%


\bibliography{refs}

\end{document}